\documentclass[sigconf]{acmart}

\AtBeginDocument{\providecommand\BibTeX{{\normalfont B\kern-0.5em{\scshape i\kern-0.25em b}\kern-0.8em\TeX}}}
\setcopyright{none}
\renewcommand\footnotetextcopyrightpermission[1]{}
\newcommand{\papertitle}{Automating Cryptographic Protocol Language Generation from Structured Specifications}
 \newcommand{\RM}{Roberto Metere}
 \newcommand{\LA}{Luca Arnaboldi}

\usepackage{units}
\usepackage{amsfonts}
\usepackage{mathtools}
\usepackage{xifthen}
\usepackage{listings}
\usepackage{xcolor}
\usepackage{graphicx}
\usepackage{subcaption}

\PassOptionsToPackage{hyphens}{url}\usepackage{hyperref} \hypersetup{
  pdfauthor={\RM, \LA},pdfkeywords={},pdftitle={\papertitle},colorlinks,                        linkcolor={red!50!black},          citecolor={blue!50!black},         urlcolor={blue!50!green!90!black}, }
\definecolor{SVGorange}{RGB}{231, 123, 0}
\definecolor{SVGgreen}{RGB}{12, 151, 0}
\definecolor{SVGblue}{RGB}{56,133,191}

\lstdefinelanguage{None}
{
  basicstyle=\ttfamily\small,
}

\lstdefinelanguage{XML}
{
  basicstyle=\ttfamily\small,
  columns=fullflexible,
  showstringspaces=false,
  morestring=[b]",
  morestring=[s]{"}{"},
  morecomment=[s]{?}{?},
  morecomment=[s]{!--}{--},
  commentstyle=\color{gray}\upshape,
  moredelim=[s][\color{black}]{>}{<},
  moredelim=[s][\color{SVGgreen}]{\ }{=},
  stringstyle=\color{SVGorange},
  identifierstyle=\bf\color{SVGblue},
}

\usepackage{semantic}

\usepackage{hypcap}

 \colorlet{colournote}{blue!80!yellow!10!white}
\definecolor{xpath-color}{rgb}{0.04,0.45,0.67}
\definecolor{xsltstr-color}{rgb}{0.8,0,0}
\definecolor{xsltattr-color}{rgb}{0.04,0.45,0.67}
\definecolor{xsltsubtag-color}{rgb}{0.67,0.45,0}

\newcommand{\observation}[1]{\todo[fancyline,size=\small,color=colournote,bordercolor=colournote]{#1}}
\newcommand{\obs}[2][]{{\ifthenelse{\isempty{#1}}{\observation{#2}}{\observation{#1: #2}}}}

\newcommand{\quotedbl}[1][]{\text{\lstinline[basicstyle=\ttfamily,stringstyle=,stringstyle=,identifierstyle=,language=]{"#1"}}}

\newcommand{\mn}[1]{\mathrm{#1}}
\newcommand{\of}[1]{\left ( #1 \right )}
\newcommand{\pbr}[1]{\left( #1 \right)}

\newcommand{\cbr}[1]{\left\{ #1 \right\}}
\newcommand{\set}[1]{\cbr{#1}}
\newcommand{\kstar}{{*}}
\newcommand{\emptystring}{\varepsilon}
\newcommand{\nestrings}{\mathcal{L}}
\newcommand{\nonce}{\mathsf{nonce}}
\newcommand{\constant}{\mathsf{const}}
\newcommand{\entity}{\mathsf{entity}}
\newcommand{\variable}{\mathsf{var}}
\newcommand{\opt}{{}^{\circ}}
\newcommand{\channelset}{\mathcal{C}}
\newcommand{\metacpmodel}{\mathcal{M}}
\newcommand{\natset}{\mathbb{N}}
\newcommand{\intset}{\mathbb{Z}}

\newcommand{\xsltrule}[2][]{\left[\mn{#2}\right]{\ifthenelse{\isempty{#1}}{}{^{#1}}}}
\newcommand{\xsltapply}[3][]{\left. \xsltrule[#1]{#2} \right|_{#3}}
\newcommand{\xsltapplyall}[2][]{\xsltapply[#1]{#2}{\forall}}
\newcommand{\xpathset}[1]{{\left< {\color{xpath-color}#1} \right>}}
\newcommand{\xsltstr}[1]{{\color{xsltstr-color}\quotedbl[#1]}}
\newcommand{\xsltattr}[2][]{#1{\color{xsltattr-color}\left[ @ #2 \right]}}
\newcommand{\xsltsubtag}[2][]{#1{\color{xsltsubtag-color}/#2}}

\newcommand{\piread}[2]{\mn{in}\of{#1, #2}}
\newcommand{\piwrite}[2]{\mn{out}\of{#1, #2}}
\newcommand{\pilet}[2]{\mn{let}\ {#1}\ \mn{in}\ {#2}} 
\newcommand{\piite}[3]{\mn{if}\ {#1}\ \mn{then}\ {#2}\ \mn{else}\ {#3}}

 \newcommand{\pigetse}[5]{\mn{get}\ #1\of{#2}\ifthenelse{\isempty{#3}}{}{\ \mn{suchthat}\ {#3}}\ \mn{in}\ {#4}\ifthenelse{\isempty{#5}}{}{\ \mn{else}\ {#5}}} 
 \newcommand{\spinew}[2]{\nu\,#1:\,#2}
\newcommand{\spisubst}[2]{\left\{ \nicefrac{#2}{#1} \right\}}
\newcommand{\spilet}[3]{\nu\,#1.{\left( \spisubst{#1}{#2}. #3 \right)}}
\newcommand{\spiread}[2][\insecurechannel]{#1\of{#2}}
\newcommand{\spiwrite}[2][\insecurechannel]{\bar{#1}{\left< #2 \right>}}

\newcommand{\thead}[1]{\bfseries #1}
 
\AtBeginDocument{\providecommand\BibTeX{{\normalfont B\kern-0.5em{\scshape i\kern-0.25em b}\kern-0.8em\TeX}}}

\begin{document}
\fancyhead{}

\title{\papertitle}

\author{\RM}
\email{roberto.metere@ncl.ac.uk}
\orcid{1234-5678-9012}
\affiliation{\institution{Newcastle University}
  \city{Newcastle upon Tyne}
  \country{UK}
}
\additionalaffiliation{\institution{The Alan Turing Institute}
  \city{London}
  \country{UK}
}

\author{\LA}
\authornote{Equally collaborative work; work partly done whilst at Newcastle University.}
\email{luca.arnaboldi@ed.ac.uk}
\orcid{0000-0002-0808-2456}
\affiliation{\institution{The University of Edinburgh}
  \city{Edinburgh}
  \country{UK}
}

\begin{abstract}
  Security of cryptographic protocols can be analysed by creating a model in a formal language and verifying the model in a tool.
All such tools focus on the last part of the analysis, verification, and the interpretation of the specification is only explained in papers.
Rather, we focus on the interpretation and modelling part by presenting a tool to aid the cryptographer throughout the process and automatically generating code in a target language.
We adopt a data-centric approach where the protocol design is stored in a structured way rather than as textual specifications.
Previous work shows how this approach facilitates the interpretation to a single language (for Tamarin) which required aftermath modifications.
By improving the expressiveness of the specification data structure we extend the tool to export to an additional formal language, ProVerif, as well as a C++ fully running implementation.
Furthermore, we extend the plugins to verify correctness in ProVerif and executability lemmas in Tamarin.
In this paper we model the Diffie-Hellman key exchange, which is traditionally used as a case study; a demo is also provided for other commonly studied protocols, Needham-Schroeder and Needham-Schroeder-Lowe.
 \end{abstract}



\maketitle

\section{Introduction}

Design and specification of cryptographic protocols are usually the first stage when creating a new protocol.
Their implementation and verification is commonly deferred to a secondary stage, and often done by a separate set of people.
At this second stage, the specification gets interpreted into a formal language able to run the protocols or verify security properties in the form of mechanised or automated theorems.
We can appreciate that such interpretations are affected by (at least) two problems: first, the language of specification may be ambiguous or contain gaps that become noticeable only at later stages, and second, proposed interpretations are difficult to reuse as they exist only on papers.
The former is a common concern when one models from specification~\cite{cremers2016improving,meier2013tamarin}.
The latter is a manual refinement from the specification language, that is often a mix of maths and natural language, to a formal language of choice, with limited semantics, that consequently may only capture a subset of the initial specification.

The model interpreting the design of a protocol is the first mathematical artefact in the process of formal verification, but the interpretation process itself is not mathematical and currently manually done by experts.
Hence, papers need to be written to convince the readers that a particular interpretation indeed captures the aspects relevant to the analysis.
As a consequence, different researchers may formalise the same specification differently, even more likely if they choose different formal languages.
The natural outcome of this process is that their output may show different results, depending on what security details are being modelled. 
Indeed, protocols proven correct by one interpretation~\cite{vigano2006automated} may be found to suffer from several vulnerabilities when formalised differently~\cite{hao2018speke}.
Nonetheless, both are valid interpretations of the same protocol.
It is therefore important to analyse the same specification from multiple interpretations that can cover security aspects more exhaustively.
Even though this concept sounds intuitive, all state-of-the-art tools exclusively focus their efforts on automating the last part of the verification process, i.e., {\em after} the model has been formalised.

We see in a structured, centralised approach for specification an effective way to tame the above mentioned difficulties.
A seminal prototype of such approach has been implemented in the tool MetaCP~\cite{arnaboldi2019poster}, which focuses on the automated modelling part of the process of formal verification, completely delegating the security proofs to external tools.
A mechanised refinement from structured specification to formal languages can offer consistency, reusability and repeatability: as if a security aspect is specified in the same manner multiple times, it will always be formalised in the same manner.
Not only does MetaCP improve a previously manual and bespoke process, but it also does so in record time and without the need of expert knowledge of multiple formal languages -- although experts may still be required to check the final results or to adjust the exported code.
We extend the core of the tool to support new plugins (ProVerif and C++) on top of previous work (Tamarin plugin); we also model {\em executability} and {\em correctness} as a first attempt to model security properties from specification\footnote{The tool is available at \url{http://metacp.eu}.}.
Section~\ref{sec:architecture} of this paper discusses in more details the architecture of MetaCP, Section~\ref{sec:workflow} its workflow, and Sections~\ref{sec:proverif}~and~\ref{sec:cpp} interpretation process.
In the latter two, we illustrate one possible interpretation from the structured specification of MetaCP to ProVerif and an interpretation to C++.
We emphasise that the tool supports for multiple interpretations across the same target language too;
so, the interpretations we propose can easily coexist with many others to both different or same languages.

 \section{Related Work}

The tools that allow to mechanise security proofs and finally perform formal security evaluation, e.g. Tamarin~\cite{meier2013tamarin}, ProVerif~\cite{blanchet2001efficient}, EasyCrypt~\cite{barthe2011computer} among others, have improved in the past decades and enjoy a wide spread use. 
However, these tools do not provide any means to relate to the whole design process, impacting the usability, reproducibility and replication of the evaluations.
As it stands, it is very difficult for the casual user, and highly time-consuming for the security professional unaware of those formal languages, to ascertain the truthfulness of a formal protocol analysis, and how it relates to the original protocol.

Witnessing the sensibility of the research community about this problem, projects such as CAPSL (Common Authentication Protocol Specification Language)~\cite{denker2000capsl}, AVISPA (Automated Validation of Internet Security Protocols)~\cite{armando2005avispa} or AVANTSSAR (Automated Validation of Trust and Security of Service-Oriented Architectures)~\cite{armando2012avantssar} have attempted to unify the verification process by presenting a single intermediate language of specification, and by automating the translation into various back-end tooling.
These proposed approaches based themselves on the assumption that most people would be familiar with their intermediate language.
Even with the integration of multiple verification options, research shows that protocols found to be secure by a tool (i.e., AVISPA)~\cite{vigano2006automated} were later found flawed when modelled manually in a different language (i.e., ProVerif)~\cite{hao2018speke}; this is (also) due to the strict semantics of the intermediate language that makes it difficult (or impossible) to capture all the specification aspects relevant to security properties.
Another approach is provided by ProScript~\cite{kobeissi2017automated} (developed as part of the tool Cryptocat -- discountinued), where authors propose a new high level language for the specification of security protocols based on Javascript.
ProScript is able to export to applied pi-calculus (without a sound translation and similar methodology to ours), whose result is automatically verifiable in ProVerif and {\em manually} in CryptoVerif.
However, its closeness to the exported language strongly limits its expressiveness, hindering its very characteristic of being more general and, thus, impractical: in fact, the task of supporting new languages from it is as difficult as it would be starting from ProVerif code itself.
Additionally, many typical specification aspects, e.g. number of bits of security parameters, cannot be even expressed in the language.

All the above demonstrate the need for simplifying the integration to more tooling in the back-end verification as the field advances.

 \section{Architecture}
\label{sec:architecture}

\begin{figure}[!ht]
  \includegraphics[width=\columnwidth]{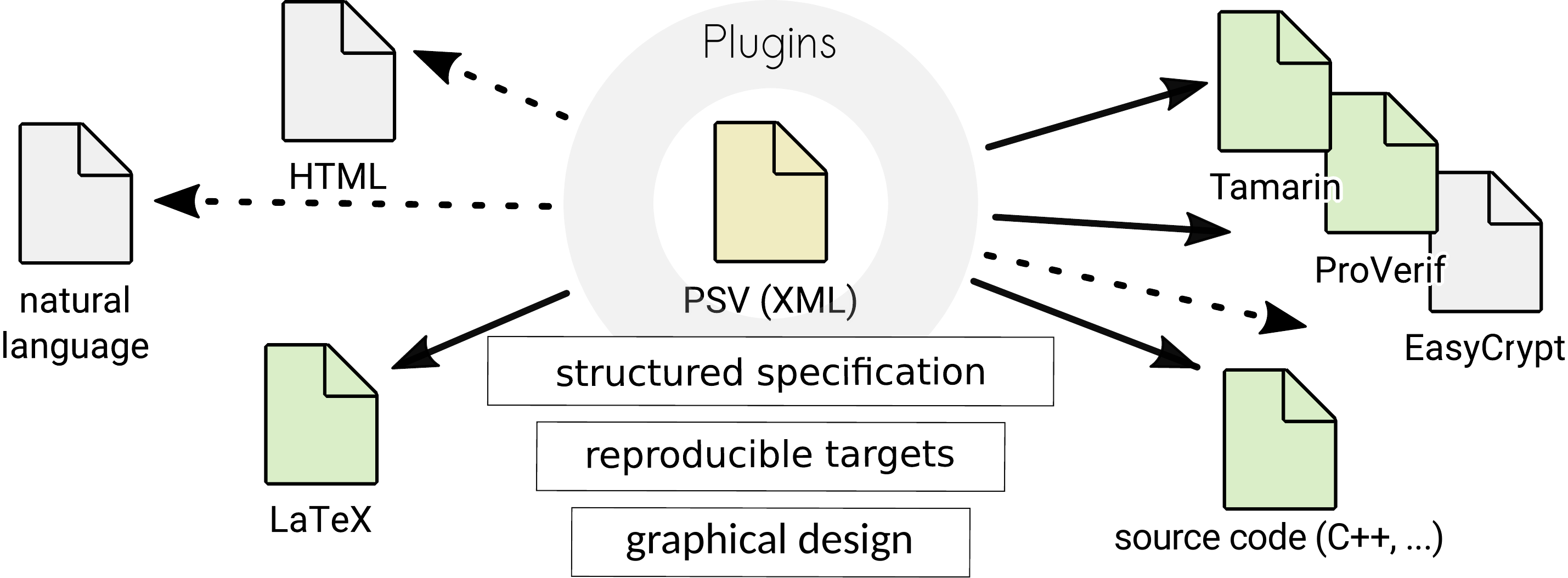}
  \caption{MetaCP supports a data-centric approach where the specification is stored as structured information (PSV). The dashed arrows point to the targets that we might attempt in future extensions.}
  \Description{PSV is a centered XML description that exports to multiple languages, like Tamarin, ProVerif, C++, LaTeX among other possible future extensions.}
  \label{fig:architecture}
\end{figure}

The crucial innovative aspect that we propose lies in a data-centric approach, where the protocol specification is stored in a structured way, as shown in Figure~\ref{fig:architecture}.
The benefits of this approach are manifold and enable for unprecedented little effort in going from the design to formal verification of security protocols.

The tool architecture is composed of three kinds of components: {\bf design}, {\bf specification}, and {\bf export}.
At the design level, an intuitive Graphical Design Editor (GDE) is provided which allows for the creation, and dragging and dropping elements that will be later saved into the specification.
The GDE is written in a modern web application framework, using ReactJs, Bootstrap, NodeJs, and Redux.
At the specification level, a data structure written in XML language is provided and is meant to collect the information required to fully describe a security protocol.
Such structure follows a minimal syntax described in Section~\ref{sec:PSV}, and its code can later be \textbf{exported} by means of a {\em plugin}.
The tool provides two plugins towards formal verification languages, one for Tamarin~\cite{arnaboldi2019poster} and one for ProVerif (presented in Section~\ref{sec:proverif}), that automatically interpret the protocol described in their syntax.
Furthermore, a third plugin exports into C++ code for which parties can truly exchange messages over the Internet and cryptographic operations are done with the Crypto++ library: we briefly illustrate it in Section~\ref{sec:cpp}.

We found it comfortable to write the new plugins in XSLT, but they can be written in any language of choice.
All components, design, specification and export, can be developed independently, and their synergy provides a tool usable to kick-start projects from the design to formal verification languages.
More details are explained in the following subsections.

\subsection{Protocol Specification and Verification Data Structure}
\label{sec:PSV}

The ability for MetaCP to automatically translate into multiple verification languages resides in its description language, denoted as Protocol Specification and Verification data structure (PSV).
A file in PSV format is effectively an XML file (although alternative formats such as JSON could be used)  whose constraints are defined in a Document Type Definition (DTD) file.
DTDs merely enforce some structure to XML files of reference without adding strong constraints to their semantics, thus not breaking the flexibility required by our approach.
Such flexibility makes PSV format suitable to be easily extended and enjoy multiple interpretations.
Our approach is sensibly different from all previous approaches, where researchers struggled to find a single semantics capable of embracing the semantics of all desired target languages~\cite{denker2000capsl,armando2005avispa,armando2012avantssar,kobeissi2017automated}.
The {\em single generic semantics} approach could work well for a few languages whose semantics were not too far apart, but would either fail, or find it very difficult, to capture the requirements of other languages.

We illustrate how our approach is suitable for being a multi-language translation tool, through a traditional example of reference in cryptography, the Diffie-Hellman key exchange (DHKE) protocol.
Once we defined the protocol in PSV, we export it to ProVerif and C++, extending previous work where a prototype of a Tamarin plugin~\cite{arnaboldi2019poster} was conceptually suggested.
We implement correctness in ProVerif and, additionally, extend the Tamarin plugin to provide {\em executability}.
The C++ plugin is able to generate compilable source code that allow parties to actually use the protocol through a real network, e.g., the Internet.

\subsubsection{Basic structure of PSV}
The tool plugins rely on the following basic structure and formalism.
Generally speaking, a party in a protocol manipulates variables whose values match a type related to some mathematical set, prepares them to be sent to the other party (or parties) through a communication channel, and elaborates the input received from the channel.
For a set $X$, we use the notation $X^\kstar$ for the Kleene closure and $\opt X$ for $X \cup \set{\bot}$ where $\bot$ is considered as {\em none}.
We start considering the set of non empty strings denoted as $\nestrings = \Sigma^\kstar \setminus \set{\emptystring}$, a set of variable modifiers $T = \set{\nonce,\constant,\entity,\variable}$, channel modifiers $\channelset_T = \set{\mathsf{insecure}, \mathsf{auth}, \mathsf{secure}}$ and {\em hints} $H$.
Hints are labels providing suggestions on the semantic interpretation of various elements.
For example, a variable may be labelled as {\em private asymmetric key}.
We do not list all hints explicitly as it is unnecessary; it will be up to the exporting plugin to interpret the hints according to the semantics of the target language.
We support probabilistic, $\mathsf{pr}$, and deterministic, $\mathsf{det}$, assignments; the former draws from a distribution over a set, the latter binds a value to an identifier.
Table~\ref{tbl:psv-syntax} shows syntactic elements in bottom-up description as sets (some are by definition mutually dependent).
\begin{table}[!ht]
\caption{Syntactic structure of the PSV.}
\Description{Syntax shows the symbol used throughout the paper, the mathematical description of the set they describe or belong to, the their corresponding syntax as XML or xPath descriptions.}
\label{tbl:psv-syntax}
\renewcommand{\arraystretch}{1}
\begin{center}
  \small
  \begin{tabular}{ccl}
    \thead{Set symbol} & \thead{Set description} & \thead{XML/xPath } \\ \hline
    $V$ & $\nestrings \times \opt T$ & $\xpathset{variable/argument}$ \\
    $P$ & $\nestrings \times \opt K$ & $\xpathset{entity}$ \\
    $K$ & $V^\kstar \times P$ & $\xpathset{knowledge}$ \\
    $\tau$ & $\nestrings \times \tau^\kstar \times \opt H$ & $\xpathset{set}$ \\
    $D_v$ & $V \times \tau \times P \times \opt H$ & $\xpathset{\xsltsubtag{/declaration\xsltattr{}}}$ \\
    $D_f$ & $\nestrings \times \opt H$ & $\xpathset{function}$ \\
    $F$ & $D_f \times \opt \pbr{V^\kstar \cup F}$ & $\xpathset{application}$ \\
    $E$ & $\nestrings \times V^\kstar \times \pbr{F, V \cup F}$ & $\xpathset{equation}$ \\
    $\channelset$ & $\nestrings \times \channelset_T$ & $\xpathset{channel}$ \\
    $A$ & $V \times \set{\mathsf{det}, \mathsf{pr}} \times \pbr{V \cup F \cup \tau}$ & $\xpathset{assignment}$ \\
    $R$ & $\opt K \times P \times A^\kstar$ & $\xpathset{finalise}$ \\
    $M$ & $\opt K^\kstar \times A^\kstar \times V^\kstar \times \channelset \times V^\kstar \times A^\kstar$ & $\xpathset{message}$ \\
    $\Pi$ & $P^\kstar \times M^\kstar \times \opt R \times \opt S$ & $\xpathset{protocol}$ \\
    $\metacpmodel$ & $\opt \tau^\kstar \times \opt D_f \times \opt D_v \times \opt E \times \Pi$ & $\xpathset{model}$ \\
  \end{tabular}
\end{center}
\end{table}

Support to security properties is still immature in MetaCP, and the tool defines only executability (Tamarin plugin) and correctness (ProVerif plugin).
Correctness is similar to executability in that it tests if the end of the protocol is reachable, but differently it also tests final conditions.
In the PSV, this notion is provided in the finalisation element, $R$.
The syntactic structure introduced in Table~\ref{tbl:psv-syntax} shows which XML tags correspond to the syntactic elements.
The full syntactic description of PSV is accessible from its DTD.
The DTD describing the structure of the PSV to specify a protocol is available here: \url{http://metacp.eu/meta-cp.dtd?v=0.1}.

\subsection{High-level description of the structure of the specification language}
\label{ap:metacp-high-level}
\begin{figure}[!ht]
  \begin{center}
    \includegraphics[width=\columnwidth]{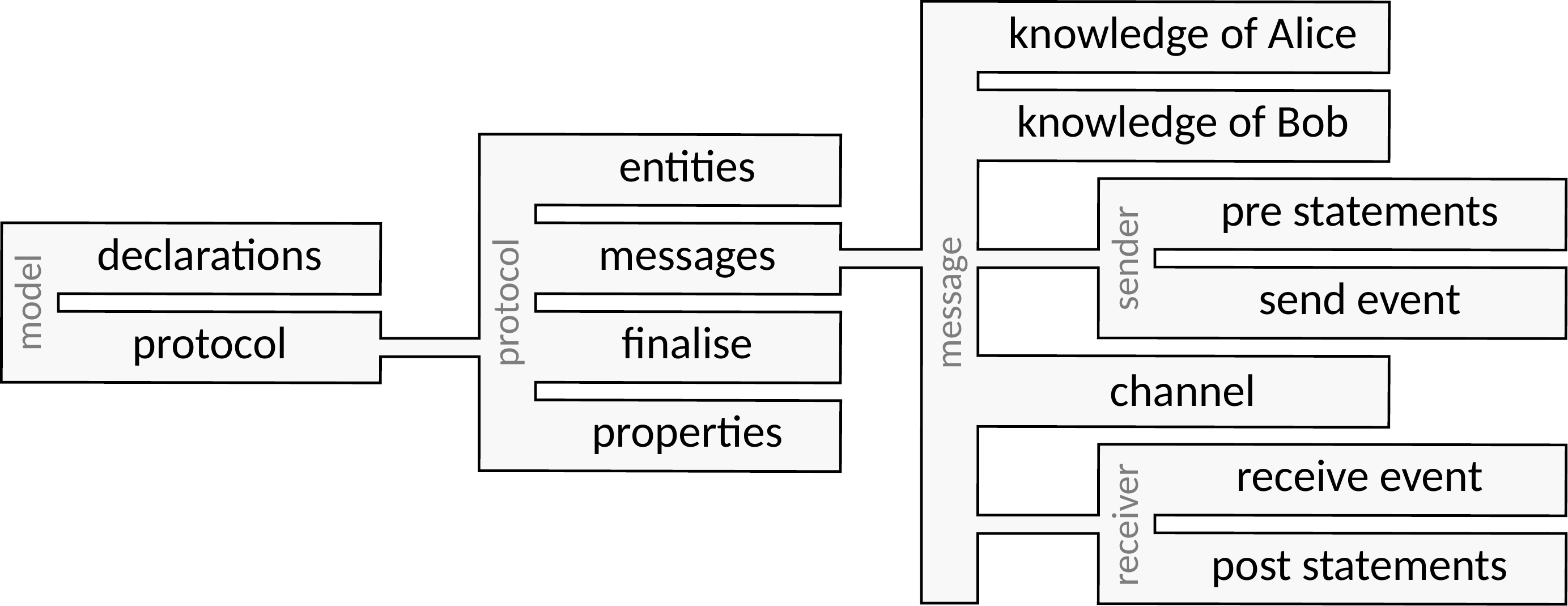}
  \end{center}
  \caption{High level description of the PSV data structure to specify protocols.}
  \Description{A model contains declarations and the protocol. The protocol contains entities, messages, a finalise section and (security) properties. Messages can be split into the knowledge of each party, a sender subpart with pre-calculated statements and a send event, a channel, and finally a receiver subpart with the receiving event and post-calculations.}
  \label{fig:psv-structure}
\end{figure}

A PSV file describes a {\em model} of a single protocol matching the syntax of Table~\ref{tbl:psv-syntax}.
Figure~\ref{fig:psv-structure} describes its general structure that includes the following sections: declarations and the protocol.

\subsubsection{Declarations.}
To allow for a type system over all the structures used within a protocol specification, e.g., typesets, variables, constants and functions, the declarations of the corresponding membership sets are mandatory beforehand.
Each subsequent declaration needs to refer to an existing set identifier.

\noindent\textbf{Typesets.}
Declarations of typesets enforce strict typing rules when constructing function applications, messages and statements.
PSV notation allows for the definition of custom sets that can be used for function declarations.
For example, we use $\mathbb{N}$ to denote the set of natural numbers and $\mathbb{Z}_p$ to define the ring of integer modulo $p$, used in modular arithmetic of the Diffie-Hellman exponentiation.
To declare such typesets, we use the markup:
\begin{lstlisting}[language=XML]
<sets>
  <set id="N">Natural numbers</set>
  <set id="Zp">Integers modulo p</set>
</sets>
\end{lstlisting}

\noindent\textbf{Variables.}
Declarations are used to preemptively specify the variables which will be used in the protocol.
In a variable declaration, one must specify its related typeset and its scope.
The scope of a variable is what entity manipulates it, assuming that the (implementation of) protocols will eventually run in different execution environments with separated memories.
\begin{lstlisting}[language=XML]
<declaration variable="x" entity="A" set="N"></declaration>
\end{lstlisting}

\noindent\textbf{Functions.}
Functions follow on from the set definitions, only existing sets may be used as an argument set (argset) of a function allowing for easy syntax checking and disallowing errors in the protocol declaration.
Whilst the PSV automatically enforces the existence of an identifier by way of tags in the DTD, it cannot check the semantic correctness of their later usage.
This problem is overcome by the consistency of identifiers enforced by the graphical interface.

A function contains not only a set of arguments but also notations and hints, which allow for the plugins to interpret the function structure efficiently.
\begin{lstlisting}[language=XML]
<function id="exp" arity="2" hint="group-exp">
    <argset set="Zp"></argset>
    <argset set="N"></argset>
    <argset set="Zp"></argset>
</function>
\end{lstlisting}
The \lstinline[language=XML]{ hint="group-exp"} attribute highlights that the function is to be interpreted as part of the group exponentiation theory, whose usage will depend on the target language.
For example, Tamarin may want to include the diffie-hellman theories with 
\begin{lstlisting}[language=None]
builtins: diffie-hellman
\end{lstlisting}
ProVerif may want to explicitly implement the commutativity
\begin{lstlisting}[language=None]
equation forall b: Zp, x: N, y: N;
  exp(exp(b,x),y) = exp(exp(b,y),x).
\end{lstlisting}
Conversely, C++ may want to use a specific library.

\subsubsection{Protocol.}
A protocol is composed of entities, messages, a final elaboration step after the messages, and finally the desired (security) properties.
The entities are the participants of the protocol that exchange messages whose directives affect their knowledge.
The final elaboration step can include statements; for example, at the end of a key exchange protocol, the parties may reconstruct the key at that stage.
Security properties that can be specified are correctness, authentication and secrecy.
In this paper, we only focus on modelling correctness (and executability for Tamarin); we reserve the study of additional security properties to future works.

The messages are structured in four parts: the {\em knowledge}, the {\em sender}, the {\em receiver} and a communication {\em channel} in their between.
The {\bf knowledge part} is per entity and lists all the known variables and constants by the entity {\em before} either sending or receiving the message.
The knowledge is beneficial to detect or restrict the designer not to use unknown structures.
The {\bf sender part} shows two sub-parts: the first can include statements required to construct the message to send, and the second is the message as it is pushed to the channel.
Similarly, the {\bf receiver part} shows two sub-parts but, in this case, they are inverted: the first is the incoming message, while the second are statements manipulating variables in the knowledge of the receiver, which has been just augmented with the received message.
We remark that the received message may not be the same as originally sent by the sender.
Any manipulation to the message can be done in the {\bf channel part}.
This structure has the benefit of allowing the designer to model different scenarios of interest.
In particular,
(i) systematically biased channels can be implemented with a function in the channel,
(ii) a man-in-the-middle may be modelled by tampering with the received message, without creating additional parties and simplifying the design of attacks, and
(iii) faults can be implemented either as empty received messages or probabilistic functions in the channel.
The above listed scenarios are merely examples, and other scenarios can benefit from this particular structure of the message.
Using the DHKE running example, we cherry-picked the first message sent in the protocol:
\begin{lstlisting}[language=XML]
<message [...] from="Alice" to="Bob">
    <knowledge entity="Alice">[...knows g]</knowledge>
    <knowledge entity="Bob">[...knows g]</knowledge>
    <pre>[...samples x, elaborates gx]</pre>
    <event type="send">[...sends gx]</event>
    <channel security="insecure"></channel>
    <event type="receive">[...receives gx]</event>
    <post></post>
</message>
\end{lstlisting}
where we replaced the details of its content with a brief summary.

\subsection{Graphical Design Editor}
\label{sec:GDE}

\begin{figure}[!ht]
  \begin{center}
    \includegraphics[width=\columnwidth]{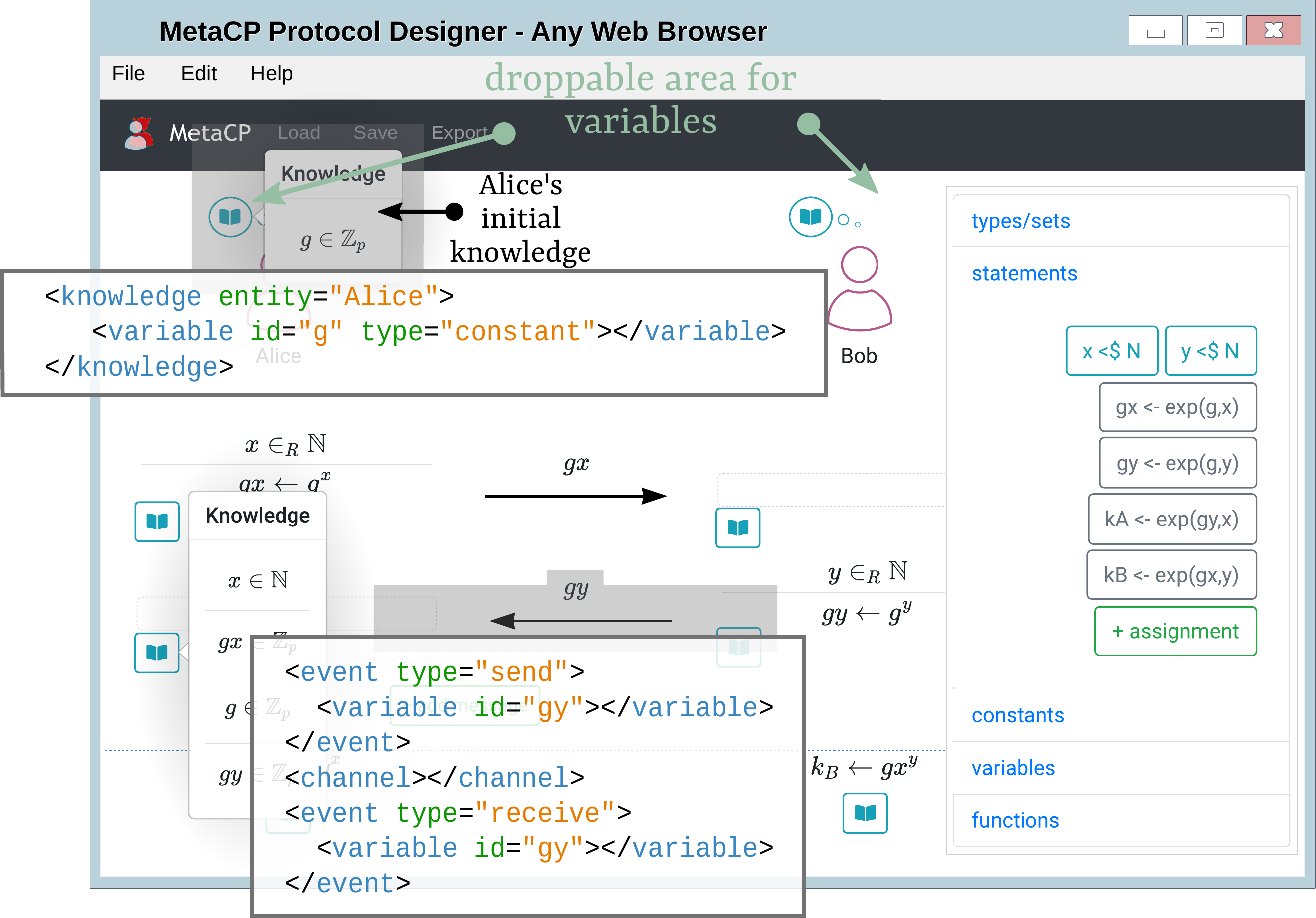}
  \end{center}
  \caption{The graphical design of MetaCP is saved as the PSV format (boxes).}
  \Description{Some parts of the graphical interface can be directly mapped to PSV.}
  \label{fig:gde2psv}
\end{figure}

XML is an intuitive language for describing a protocol in a specification, as its format is purposely easy to be manipulated by both humans and machines.
Additionally, MetaCP is equipped with a Graphical Design Editor (GDE).
The GDE aids the user with the design of the protocol rather than focusing on the formalisation part, i.e., the PSV.
The GDE mimics the standard drawing process most familiar to any protocol designer, and it lets the user specify variables, functions and message flow.
It does so through a smooth drag and drop design, making it easy to piece together the protocol.
The GDE is intended to guide a user through the coherent definition of the PSV, automatically providing the following relationships in the data structure:
first, the knowledge is automatically augmented as the protocol is constructed, and second,
the GDE will enforce correct typing across the protocol and functions.
The ability to store further relations and information about the protocol is a significant aid that modern frameworks for programming languages usually incorporate as the basics - but currently is not supported in any existing automated protocol verification tool.
Once a desired protocol is drawn out, it can be saved as PSV.
Figure~\ref{fig:gde2psv} highlights how some parts of the design reflect to the structure in PSV format.

\subsection{Exporting Plugins}

A plugin provides a fully automated protocol-agnostic interpreter from PSV code to the desired semantics of the target language. 
We remark that our plugins are examples of interpretation of the target semantics: additional plugins targeting the same language are allowed. 

The combination of the benefits of the GDE and the exporting plugins can sensibly improve the experience of protocol designers, even if they are an expert in a specific language.
To the best of our knowledge, the languages used in formal verification for protocols do not enjoy frameworks for design or editing.
In addition, many tools work (or can work) with untyped variables and constants.
So in comparison to other languages, their source code is more prone to subtle and hard-to-spot bugs that can influence the consistency of the specification.
Imagine simply asking for the confidentiality of a never-used variable - due to a typo - it can verify {\em correctly} as unknown to any attacker.

Plugins can be called in the GDE directly, as well as natively, e.g., as scripts in a shell, once the PSV is available.
The architecture of MetaCP is such that all the components, PSV (with DTD), GDE and exporting plugins, are independent.
So when a plugin is called in the GDE, it automatically and transparently generates a PSV as input to the plugin.

\section{Workflow}
\label{sec:workflow}

The aim of MetaCP is to facilitate the design and modelling process up to the formal verification (excluded) of a cryptographic protocol.
With this goal in mind, the ideal workflow of its user can be summarised by the following points:
\begin{enumerate}
  \item Design the protocol with the aid of the graphical design editor of MetaCP.
  \item Save the design to PSV format, that ideally specifies the protocol.
  \item Export the PSV to any target language or format, e.g., pi-calculus.
  \item Optionally run the formal verification tool, e.g., ProVerif, to formally verify the protocol model or execute the protocol code exported (discussed for completeness, although it sits outside MetaCP).
\end{enumerate}
Only the first and the last step are interactive, as both saving and exporting are automated.
Nevertheless, the user can intervene and manually modify the result of any step of the process.
After saving to PSV, the user might enrich the PSV with additional information that are not supported by the GDE, e.g., with additional properties.
Similarly, after exporting to pi-calculus, the user might modify the model if required to verify later in ProVerif.
Modifying the PSV is as easy as modifying an XML file.
Differently, modifying the exported protocol requires expertise in the target format or language.
The application of our exporting plugin (e.g., PSV-to-ProVerif) can be as {\em good} (or as {\em wrong}) as the manual modelling task (e.g., English-to-ProVerif); while both need to show convincing arguments, only the former is mechanised and can be consistently reused.

\subsection{Graphical design editor}
To show the design process, we provide some details of the MetaCP graphical design editor, introduced in Section~\ref{sec:GDE}.
The MetaCP GDE is composed by the following macro-blocks: a section with two parties, a toolbox, the exchanged messages and the final operations.

\textbf{Parties.}
Currently, the GDE supports two parties, Alice and Bob, each with their knowledge, as shown by Figure~\ref{fig:gde2psv}.
The user can drop variables to the knowledge of either party, determining  their initial knowledge, i.e., what they know before running the protocol.

\begin{figure}[!ht]
  \caption{Toolbox in the GDE.}
  \Description{A toolbox contains several sub parts, for types, statements, constants, variables and functions. For example, functions show both function definitions and function applications (with specified arguments).}
  \label{fig:gde-toolbox}
  \begin{center}
    \includegraphics[width=\columnwidth]{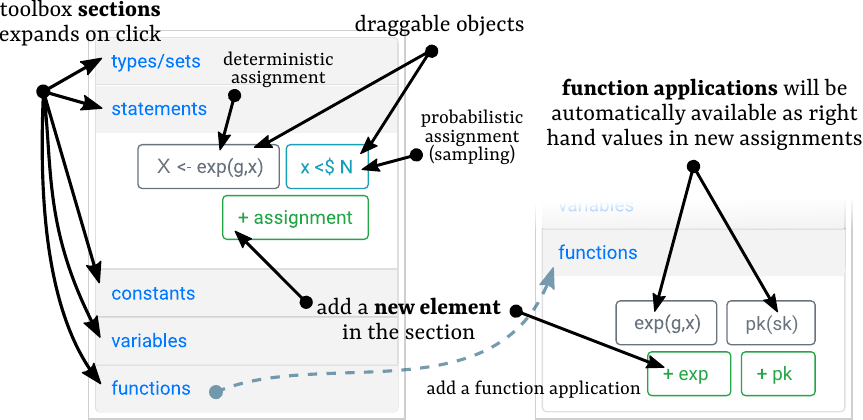}
  \end{center}
\end{figure}

\textbf{Toolbox.}
The toolbox is illustrated in Figure~\ref{fig:gde-toolbox}.
The toolbox is split in sections for handing sets, functions, constants, variables and statements.
They contain buttons to add new elements.
Functions themselves cannot be directly dragged out of the toolbox, and they require the user to create an application, i.e., an instance that specifies arguments to the function.
A type match helps the user to avoid incorrect function applications or statements.
Once new objects are created, they can be dragged and dropped to target boxes external to the toolbox.

\begin{figure}[!ht]
  \caption{Message structure in the GDE.}
  \Description{The structure of a message is in three parts. A box called pre where calculations are done before sending the message. A box with the content of the message. And finally a box called post with calculations that are done after the message has been received.}
  \label{fig:gde-messages}
  \begin{center}
    \includegraphics[width=.9\columnwidth]{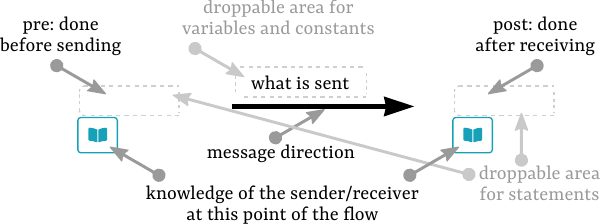}
  \end{center}
\end{figure}

\textbf{Messages.}
The structure of the messages exchanged by Alice and Bob is illustrated in Figure~\ref{fig:gde-messages}.
An arrow shows the direction of the message from the sender to the receiver.
The user can drop statements from the toolbox to the {\em pre} and {\em post} boxes that are computations made by the parties before and after the message has been exchanged.
A box with knowledge is automatically populated according to the initial knowledge and the computations made by the parties in the message.
Additionally, the user can drop variables and constants to the {\em event} box above the arrow, representing what is being sent by the sender.

\textbf{Final operations.}
Final operations are statements computed (offline) by either of the parties after all message have been exchanged.
For example in the case of a key exchange protocol, they may contain the final elaboration of the exchanged key done by each party.

\subsection{Diffie-Hellman key exchange}
\label{sec:dhke-model}
As a case study, we show some details of the workflow to successfully design and formally verify the Diffie-Hellman key exchange (DHKE) protocol, a traditional protocol of reference in cryptography.
This case study saves to PSV and then exports to ProVerif as intended, i.e., no manual intervention is required.

The workflow can be summarised by the following steps:
\begin{itemize}
  \item Declare the sets, constants and  variables that will be used across the protocol.
  These can be found in the toolbox (see Figure~\ref{fig:gde-toolbox}) and are applied as follows:
  \begin{itemize}
    \item {\em TypeSets }, {\tt N}, for the exponents, and {\tt Zp}, for the group.
    \item {\em Constants},  {\tt g}, the group generator is a constant as it is a pre-shared knowledge before engaging with the protocol.
    \item {\em Variables},  {\tt x, y: N} for fresh and secret exponents, {\tt X, Y: Zp} for the messages, and {\tt kA, kB: Zp} for the key they share at the end.
  \end{itemize}
  \item Create the function applications that will be assigned to variables.
  \begin{itemize}
    \item {\em Secret exponent sampled by Alice}, {\tt x <\$ N}
    \item {\em Secret exponent sampled by Bob  }, {\tt y <\$ N}
    \item {\em Message from Alice}, {\tt X <- exp(g,x)}
    \item {\em Message from Bob  }, {\tt Y <- exp(g,y)}
    \item {\em Shared key reconstructed by Alice}, {\tt kA <- exp(Y,x)}
    \item {\em Shared key reconstructed by Bob  }, {\tt kB <- exp(X,y)}
  \end{itemize}
  \item Add two (empty) messages the first from Alice to Bob and the last from Bob to Alice, that will have the structure shown in Figure~\ref{fig:gde-messages}.
  \item Drag statements to the {\tt pre} boxes of the messages that calculate the messages to send, then drag the variables and constants to define the message exchanged.
  In particular, the first message will be:
  \begin{itemize}
      \item {\em Alice calculates} {\tt x <\$ N} and {X <- exp(g, x)}
      \item {\em Alice sends out} {X}
  \end{itemize}
  Analogously, the second message will be:
  \begin{itemize}
      \item {\em Bob   calculates} {\tt y <\$ N} and {Y <- exp(g, y)}
      \item {\em Bob   sends out} {Y}
  \end{itemize}
  \item Fill the finalise boxes, as seen at the bottom of Figure~\ref{fig:gde2psv}.
  \begin{itemize}
      \item {\em Alice constructs the key} {\tt kA <- exp(Y,x)}
      \item {\em Bob   constructs the key} {\tt kB <- exp(X,y)}
  \end{itemize}
  \item Finally, drag variables to the initial knowledge of the parties.
  All the following knowledge bubbles will be automatically populated.
\end{itemize}
Once the design of the protocol is completed, we can save its specification to file that can be later interpreted by plugins.

 \section{Exporting plugins}
\label{sec:exporting-plugins}

The PSV is refined into various target languages through the use of plugins -- 
these will be described in the following sections.
Since the target languages we discuss in this section are different, we provide a brief discussion of main high-level differences in Section~\ref{sec:plugins}.

\subsection{Exporting to ProVerif}
\label{sec:proverif}

While the PSV is a structured container of the specification of the protocol, an interpreting plugin confers semantics to that specification from the point of view of their target language.
Hence, a plugin can be seen as the effort of applying the semantics of the target language to the structure of the source PSV.
To do that, the plugin translates the PSV into the target language grammar. 
For this paper we illustrate an example of exporting to ProVerif, for reference the syntax of ProVerif is illustrated in Figure~\ref{fig:proverif-syntax}.
\begin{figure}[!ht]
  \small
  \begin{tabular}{lll}
    $E$ & ::= & \bf Expressions \\
        & $M$ & term (variables, names, constructors) \\
        & $f(E,E,\dots,E)$ & function application \\
    $P$ & ::= & \bf Processes \\
        & $0$ & no operations \\
        & $\piwrite{\bar{M}}{M}$ & write $M$ to channel $\bar{M}$ \\
        & $\piread{\bar{M}}{x:\tau}$ & bind $x$ (of type $\tau$) reading from $\bar{M}$ \\
        & $P|P$ & parallel composition \\
        & $!P$ & (infinite) replication \\
        & $\spinew{x}{\tau}$, P & restriction (probabilistic assignment) \\
        & $\pilet{x = E}{P}$ & expression evaluation \\
        & $\piite{M}{P}{P}$ & conditional \\
  \end{tabular}
  \caption{Extract of the core syntax of ProVerif~\cite{blanchet2016modeling}.}
  \Description{The core syntax of ProVerif of expressions and processes. Processes are used to describe algorithms and messages exchanged between participants.}
  \label{fig:proverif-syntax}
\end{figure}

\begin{figure}[htbp]
  \begin{center}
    \includegraphics[width=\columnwidth]{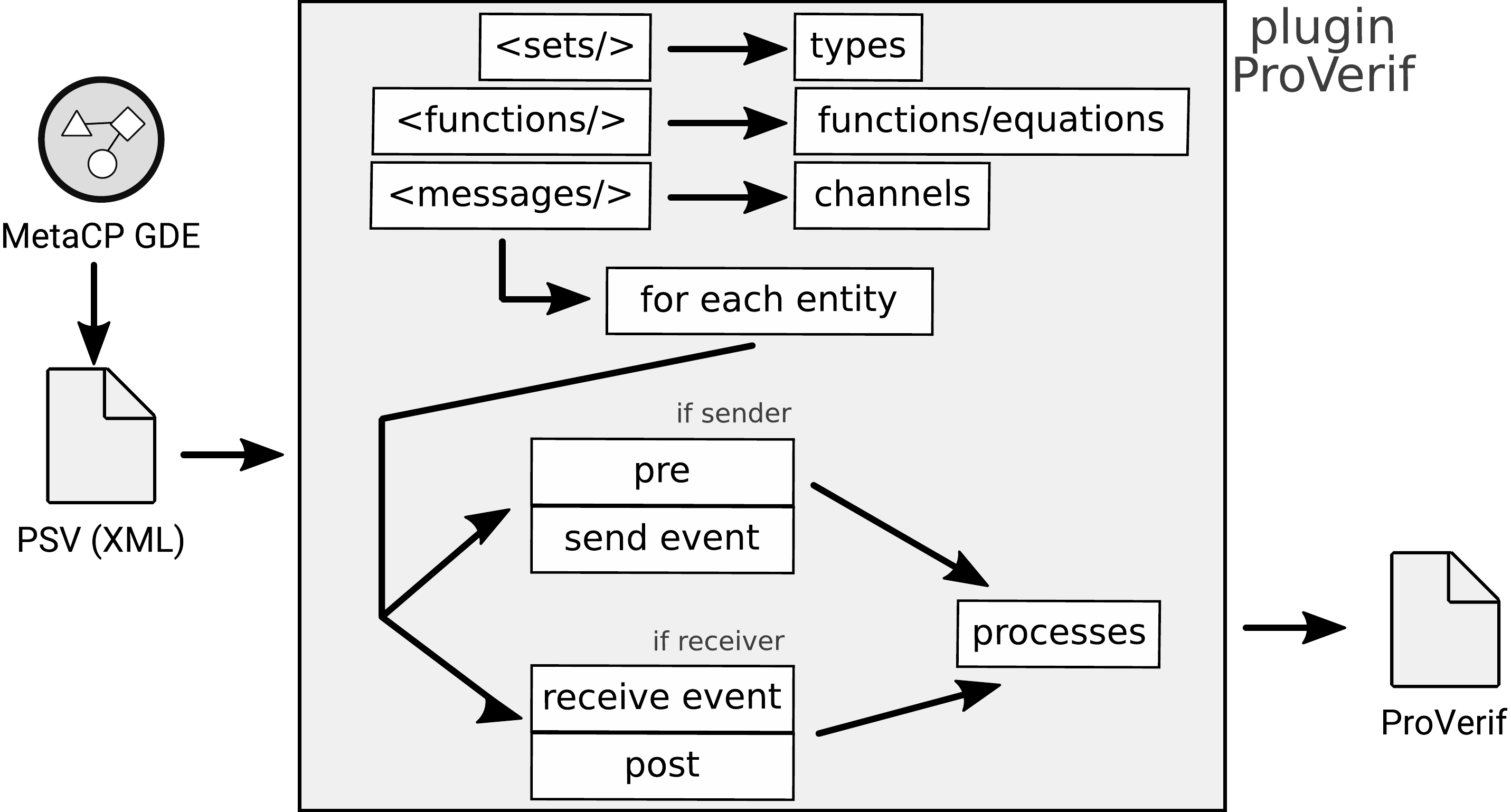}
  \end{center}
  \caption{The interpretation process automatised in the ProVerif plugin of MetaCP.}
  \label{fig:proverif-plugin-methodology}
  \Description{Starting with the graphical interface, you can save the protocol as PSV; then the plugin interprets the tags generating code in the ProVerif.}
\end{figure}

If the PSV and target language were two languages with their own semantics, some sort of bisimulation certifying that semantics are preserved in the translation would be expected.
Conversely in our case, the best we can do is to illustrate how the methodology of our ProVerif plugin does not introduce errors in the target code upon certain conditions.
Arguably, the most convenient way to describe the {\em validity} of the interpretation of a plugin is by demonstrating how the PSV structure uniquely maps to the target syntax to model the protocol, similarly to a refinement.
The interpretation methodology is illustrated in Figure~\ref{fig:proverif-plugin-methodology} and can be summarised in the following points:
(i) in a first step it handles declarations and descriptions of types, entities, functions (including constants) and channels,
(ii) then it creates the processes by entity, extracting from the messages the relevant parts, and
(iii) it creates a process describing the protocol run for infinite repetitions of the two entity-related processes.

The semantics that the DTD confers to the PSV is very intuitive and is meant to be interpreted directly.
The plugin rules complement them by applying the target language semantics to the syntax of PSV.
We only describe the rules in charge of interpreting a by-entity {\em process} in pi-calculus, as they where the least obvious and probably the most interesting, in Figure~\ref{fig:proverif-plugin-rules-easy} and Figure~\ref{fig:proverif-plugin-rules}.
We note that our plugin converts the protocol model through a generic interpretation of the rules, agnostic of security properties to verify.
Whilst this is enough for our case study, it may not be the case for other properties in different protocols.
In such cases, additional plugins may convert the model in a way that is specific for the security properties being verified.
yThe key to read those rule is as follows: at the conclusion (bottom or bottom-right) we have the grammar of the applied pi-calculus, which is inferred by the parts above its line or at the left of the corresponding inline symbol $|-$, while the lines above are the interpretation reading from the PSV whose notation uses xPath directives.
For a comprehensive explanation of the applied pi-calculus grammar, syntax and semantics, we refer to Abadi et al.~\cite{abadi2017applied}; similarly for the xPath directives, we refer to Clark et al.~\cite{clark1999xml}.
\begin{figure}[htbp]
    \begin{equation*}
    \renewcommand\arraystretch{3}
      \resizebox{\columnwidth}{!}{$\begin{array}{c}
        \inference[{[variable|argument]\textsuperscript{m}}: ]{
              v \in \xpathset{variable|argument}
            & m = \bot}
          {x \gets \xsltattr[v]{id} |- {x}} \\
        \inference[{[variable|argument]\textsuperscript{m}}: ]{
              v \in \xpathset{variable|argument}
            & m = \xsltstr{typed}}
          {
            \inference{
              \tau \gets \xpathset{\xsltsubtag{/declaration\xsltattr{variable = \xsltattr[v]{id}}}\xsltsubtag{\xsltattr{set}}}
            \\ x \gets \xsltattr[v]{id}
            }
            {x: \tau}
          } \\
        \inference[{[application]}: ]{
              a \in \xpathset{application}
          }
          {
            \inference{
              f \gets \xsltattr[a]{function}
            & l \gets \xsltattr[f]{arity}
            \\ M_1, M_2, \dots, M_l \gets \xsltapplyall[\xsltstr{typed}]{application|argument}
            }
            {f\of{M_1, M_2, \dots, M_l}}
          }
      \end{array}$}
    \end{equation*}
  \caption{Rules applied by the ProVerif plugin in MetaCP for \textbf{variables}, {\bf arguments} and function {\bf applications}.
  }
  \Description{Transformation rules for variables, arguments and applications interpreting the specification.}
  \label{fig:proverif-plugin-rules-easy}
\end{figure}

Additionally, we use the notation explained as follows.
We refer to the (ordered) sets of elements generated by an application of an xPath directive $d$ within angle brackets, i.e. $\xpathset{d}$.
If the result set is a singleton, we also refer to the single element with the same notation.
Some rules, names inside square brackets, are parametric, the parameter passed to them is superscripted after their name, so the notation [r]$^p$ is for the rule named ``r'' with parameter $p$.
Parameters do not have corresponding attributes in PSV, but are different interpretations of the same tags from the plugin.

\begin{figure}[t!]
    \begin{equation*}
        \renewcommand\arraystretch{3}
      \resizebox{.9\columnwidth}{!}{$\begin{array}{c}
        \inference[{[entity]}: ]{e \in \xpathset{entity}}
          {P \gets \xsltapplyall[e]{message}.0 |- P} \\
        \inference[{[message]}\textsuperscript{e}: ]{m \in \xpathset{message} & \xsltattr[m]{from} = e}
          {P \gets \xsltapplyall{pre}.\xsltapply{event}{\forall e \in \xpathset{event[1]}} |- P} \\
        \inference[{[message]}\textsuperscript{e}: ]{m \in \xpathset{message} & \xsltattr[m]{to} = e}
          {P \gets \xsltapply{event}{\forall p \in \xpathset{event[2]}}.\xsltapplyall{post} |- P} \\
        \inference[{[pre],[post]}: ]
          {p \in \xpathset{pre|post}}
          {P \gets \xsltapplyall{assignment} |- P} \\
        \inference[{[assignment]}: ]{
              a \in \xpathset{assignment}
            & \xsltattr[a]{type} = \xsltstr{probabilistic}
          }
          {
            \inference{
                \inference
                  {n \gets \xsltattr[a]{variable}}
                  {\tau \gets \xpathset{\xsltsubtag{/declaration\xsltattr{variable = n}}}}
            }
            {\spinew{n}{\tau}}
          } \\
        \inference[{[assignment]}: ]{
              a \in \xpathset{assignment}
            & \xsltattr[a]{type} = \xsltstr{deterministic}
          }
          {
            \inference{
                  x \gets \xsltattr[a]{variable}
                & M \gets \xsltapplyall{application}
            }
            {\spilet{x}{M}{\bullet}}
          } \\
        \inference[{[event]}: ]{
              e \in \xpathset{event}
            & \xsltattr[e]{type} = \xsltstr{send}
          }
          {v \gets \xsltapplyall[\bot]{variable} |- \spiwrite{v}} \\
        \inference[{[event]}: ]{
              e \in \xpathset{event}
            & \xsltattr[e]{type} = \xsltstr{receive}
          }
          {v \gets \xsltapplyall[\xsltstr{typed}]{variable} |- \spiread{v}} \\
      \end{array}$}
    \end{equation*}
  \caption{Rules applied by the ProVerif plugin in MetaCP for messages in the protocol along with depending rules. Generic insecure channel denoted as $\varepsilon$.}
  \Description{Transformation rules for entities, messages, assignments, pre and post blocks, and events interpreting the specification.}
  \label{fig:proverif-plugin-rules}
\end{figure}
To read attributes from tags, we use square brackets notation traditional in xPath, so we denote the attribute {\tt type} from the tag in $e$ as $\xsltattr[e]{type}$.
Unlikely other common rules, they have to be explicitly called.
The notation we use to apply a rule to all elements of a set of elements is a vertical bar with the application domain as subscript, e.g. to apply the rule [r] to all elements in $\xpathset{d}$, we write $\xsltapply{r}{\forall e \in \xpathset{d}}$.

As a short notation, if the rule to apply has the same name as the xPath directive of the set, we omit it leaving only the $\forall$ symbol, e.g.
$ \xsltapplyall{r}$ is short for $\xsltapply{r}{\forall e \in \xpathset{r}}. $
Finally, we shorten the call of two rules applying to set of diverse elements, e.g. el1 el2 with the vertical bar $|$, e.g. $\xsltrule{el1|el2}$ will apply to elements whose tag is either el1 or el2 in the order they appear in the application domain.
So for example, the PSV assignment
\begin{lstlisting}[language=XML]
<assignment variable="x" type="probabilistic"></assignment>
\end{lstlisting}
is transformed by the rule [assignment] in Figure~\ref{fig:proverif-plugin-rules} to $\pi$-calculus as $\spinew{x}{\natset}$ (see Figure~\ref{fig:proverif-syntax}), where $\natset$ is the typeset specified in the declaration of $x$, ultimately written as {\tt new n:N;} in ProVerif.

As the reader may already have noticed, the rules in Figure~\ref{fig:proverif-plugin-rules-easy} and Figure~\ref{fig:proverif-plugin-rules} rely on some assumed relationships between the elements in the PSV.
These relationships cannot be enforced by the DTD: for example, the rule $\xsltrule[\xsltstr{typed}]{variable}$ assumes that the type will be actually found in the declarations.
The DTD can only guarantee that the variable specified appears as an {\em identifier} in the past, but it cannot guarantee that the identifier was actually defined for the desired element.
By designing the protocol with the GDE, MetaCP is able to generate a PSV where these relationships are always valid.
A further area of exploration would be to enforce such relationships in the PSV itself: we reserve this investigation to future extensions, perhaps upgrading the DTD to the more powerful XML Schema Definition~\cite{thompson2009w3c}.
As introduced before, to overcome the limitations of the DTD, the GDE currently confers a type system to the functions, variables and statements to be respected across the whole protocol, and manages the knowledge automatically at each step of the protocol.
In particular, the extra properties provided by using the GDE are that all messages are exchanged between intended parties, and all statements can refer only to corresponding pre-declared sets, variables and functions, according to the knowledge of the party at that specific point of the protocol.
 \subsection{Exporting to C++}
\label{sec:cpp}

To provide a broader evaluation of the expressivity of the PSV in its current development (see the DTD for details), we created a plugin that targets C++ and allows for two parties to truly run the protocol over the Internet.
Even if the plugin is agnostic of the particular protocol under design, we tested it only for our use case, the Diffie-Hellman key exchange.
Currently, the PSV does {\em not} contain as many implementation details as a programming language, i.e. C++, so it is necessary for the plugin to make some assumptions: the protocol is a two-party protocol, and two sets $\natset$ and $\intset_p$ are present.
The plugin looks for group exponentiation operations, in particular:
\begin{itemize}
  \item The group exponentiation constant of type $\intset_p$ is the generator, and the constant of type $\natset$ is considered as the modulo $p$. We remark that such values are globally declared, so that entity-scoped values will not be treated the same way.
  \item If a function for modular exponentiation ({\tt hint}) with signature $\exp: \intset_p -> \natset -> \intset_p$ is found, its implementation gets filled with modular exponentiation with modulo $p$, where $p$ references the global value $p$.
\end{itemize}
The implementation of cryptographic functions is borrowed from the library Crypto++~\cite{cryptopp}, the implementation of the network operations is borrowed from the library Asio~\cite{asiocpp}.
Those are examples of {\em interpretation} details that are allowed by the structured nature of the PSV; although in the future, such details may be very well part of its definition - to strictly specify that use of a particular routine is mandated by the specification.

\begin{figure}[!ht]
  \begin{center}
    \includegraphics[width=\columnwidth]{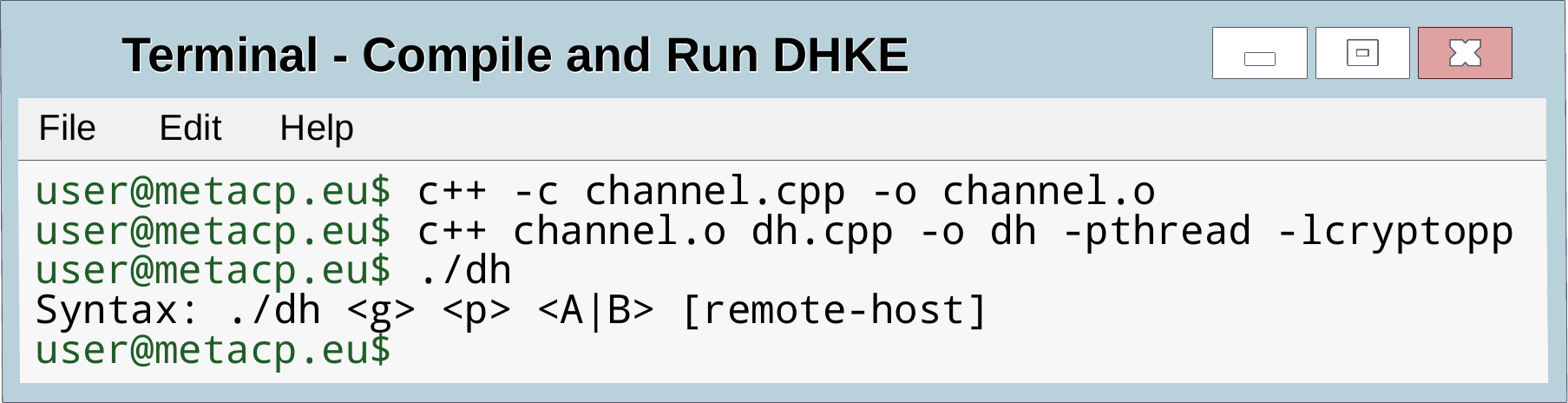}
  \end{center}
  \caption{Compiling the source code automatically generated by the C++ plugin of MetaCP. For simplicity, we assume we're using a *nix machine.}
  \Description{The generated code can be compiled straight away and creates a single executable that can be used by either of the parties.}
  \label{fig:cpp-compile}
\end{figure}

\begin{figure}[!ht]
  \begin{center}
  \noindent\includegraphics[width=\columnwidth]{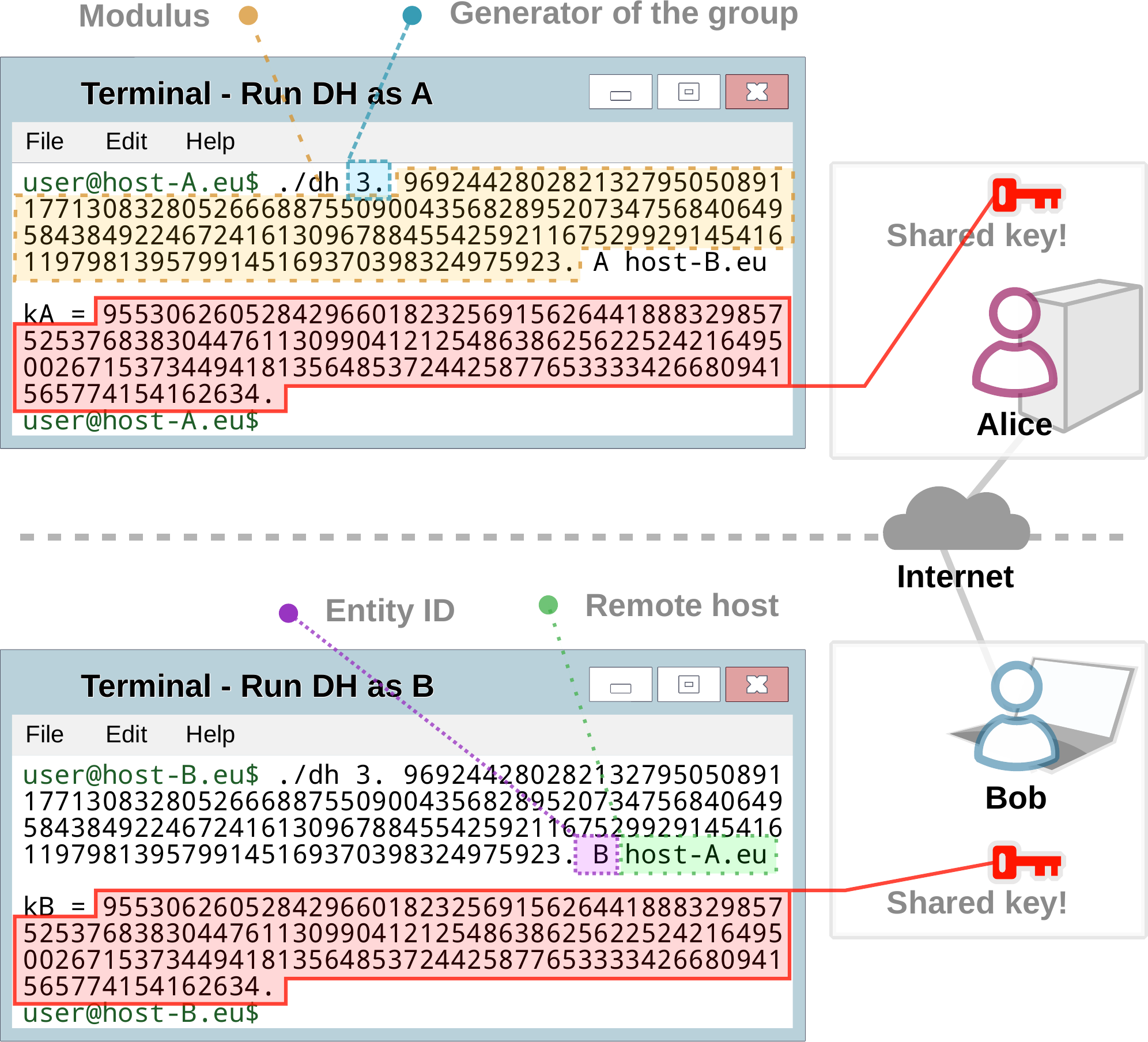}
  \end{center}
  \caption{An actual run of the Diffie-Hellman key exchange protocol, directly from the design. As we see the two hosts are able to exchange a secret key {\em correctly}.}
  \Description{Alice and Bob communicate through the Internet after compiling the automatically generated code. At the end they can share a secret key.}
  \label{fig:cpp-exec}
\end{figure}

We do not go in further details of the C++ plugin, as they are analogous to the process explained in Section~\ref{sec:proverif} for the ProVerif plugin.
Rather, we go into the details of how to compile and run the code, to appreciate the actual usability of the automatically generated implementation.
Additionally to the above mentioned libraries, the plugin relies on an external open-source class (available at:  \url{https://github.com/nitrogl/snippets/}); one needs to download and compile {\tt C++/net/channel.cpp} and {\tt C++/net/channel.h} to easily map send and receive operations.
Once all dependencies are installed, assuming that the automatically generated C++ code is saved as {\tt dh.cpp}, compilation is straightforward, see Figure~\ref{fig:cpp-compile}.
We notice that the generator $g$ and the prime number $p$ defining the modulus of the group set $\intset_p$ are publicly known by the entities before they run the protocol.
Hence, they are asked as arguments.
In the example in Figure~\ref{fig:cpp-exec} we used the following:
\begin{equation*}
  \begin{split}
    g = &\ 3 \\
    p = &\ 9692442802821327950508911771308328052666887550900435 \\
        &\ 6828952073475684064958438492246724161309678845542592 \\
        &\ 11675299291454161197981395799145169370398324975923 \\
  \end{split}
\end{equation*}where $p$ is a 512-bit prime number and ``512'' is the security parameter specified in the PSV.

To enact the algorithm as either Alice or Bob, an additional argument is required that matches the identifier of the entity used in the PSV.
Finally, the remote host to send messages to can be specified as {\em optional} argument - {\tt localhost} is used if omitted.
The execution runs between machines over the Internet\footnote{As a network detail, for two (non-isolated) machines in the same network, no further operations are usually required. Conversely in the Internet, commonly hosts are behind a router. In this case, server ports need to be opened either manually (port forwarding or triggering) or automatically (UPnP, DMZ).}, as shown in Figure~\ref{fig:cpp-exec}.

We remark that the Diffie-Hellman key exchange protocol is known to suffer from man-in-the-middle attacks; so it is not of particular interest in the real-world.
Traditionally though, the DHKE is used as a reference protocol comparing with related work.
Its attacks can be easily shown by ProVerif and Tamarin; a few manual tweaks are required with the MetaCP auto-generated scripts.
The C++ plugin can be used to generate instances of the other two available examples, the Needham-Schroeder and the Needham-Schroeder-Lowe protocols.

 \subsection{Protocol refinement using plugins for ProVerif and C++}
\label{sec:plugins}
\begin{figure}[!ht]
  \caption{The plugins refining the PSV to C++ and ProVerif.}
  \Description{We show how some blocks in the PSV description (as XML) are mapped to ProVerif or C++ by the corresponding plugin.}
  \label{fig:psv-proverif-cpp}
  \begin{center}
    \includegraphics[width=\columnwidth]{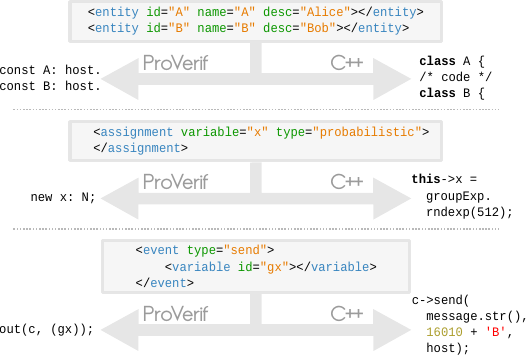}
  \end{center}
\end{figure}

In the following, we provided a quick comparison between the specification in PSV format and the output of plugins to appreciate structure similarities and dissimilarities.
In particular, Figure~\ref{fig:psv-proverif-cpp} shows different PSV elements and roughly the corresponding code output by two plugins, ProVerif and C++.
The applied pi-calculus (ProVerif) is a symbolic description of protocols, so for the most its corresponding PSV code is straightforward.
Though, it sometimes needs to navigate to other tags, i.e. the channel name {\tt c} in {\tt out(c, (gx));} in the bottom conversion is taken from outside of the tag.
Conversely, C++ is not symbolic and needs to know what exact algorithm to run.
This can be seen in the conversion of the probabilistic assignment, where a specific {\tt groupExp.rndexp} function is called; while in ProVerif the sampling distribution is abstract and (uniformly) draws elements from the corresponding declaration typeset.
In the same conversion, the parameter {\tt 512} is a {\em security parameter} that is actually specified in the PSV, but outside of the assignment.
 \section{Security properties}
\label{sec:security-properties}

Our first attempt to model security properties models correctness in ProVerif and executability in the case of Tamarin.
They are similar properties: the former checks if a {\em final} condition is met after a honest execution of the protocol (e.g. for a key agreement protocol, all parties must share the same key), the latter simply checks whether all the rules (Tamarin uses multiset rewriting rules to specify the protocol) describing the protocol can run in the desired order.
Thus, correctness may fail even if executability holds.
While correctness is a traditional security property, executability can be considered an extremely weak property~\cite{basin2018formal} that only shows that the model can run to completion; so, it is mostly used as a self-check that the model in Tamarin is not affected by (major) errors.

\subsection{Executability}
\label{sec:executability}

Executability is a security property that does not require changes in the structure of the protocol described as PSV, as the definition of the protocol already contains enough information to formalise the property.
In Tamarin, executability is generated by the plugin in the following way.
\begin{itemize}
  \item Per message, two rules (sub-procedures) are generated, one to send and one to receive.
  \item Each rule contains a unique {\em action} representing each message being sent or received, e.g. {\tt Send\_m1(A,B,m1)} for the first message {\tt m1} sent from {\tt A} to {\tt B}; the order of when such actions happen are recorded in the {\em execution trace} of the reasoning core of Tamarin.
  \item We write a lemma requiring that the above actions happen {\em in temporal order}:
  \begin{lstlisting}[language=None]
lemma executable_protocol: exists-trace
  "Ex A B m1 #a #b m2 #c #d.
       Send_m1(A, B, m1) @ #a
  & Receive_m1(B, A, m1) @ #b
  &    Send_m2(B, A, m2) @ #c
  & Receive_m2(A, B, m2) @ #d
  & #a < #b & #b < #c & #c < #d"
  \end{lstlisting}
  where {\tt A,B} are parties, {\tt m1,m2} are variables, {\tt \#a}, {\tt \#b}, {\tt \#c}, {\tt \#d} are temporal (ordered) values and {\tt Send\_m1}, {\tt Receive\_m1}, {\tt Send\_m2}, {\tt Receive\_m2} are actions in the rules.
\end{itemize}
The above lemma translates to checking whether a trace exists in which {\tt A} has sent {\tt B} a message {\tt m1} at time {\tt a} and {\tt B} has received the same message at a different time {\tt b}.
Executability is captured by a trace where all rules are in the right order, i.e., {\tt \#a} < {\tt \#b} < {\tt \#c} < {\tt \#d}.

\subsection{Correctness of key-exchange protocols}
\label{sec:correctness}
We now describe how we modelled correctness in the specification language.
Correctness is modelled by mean of the tag $\xpathset{correctness}$.
Inside of it, one can specify a relation that evaluates to boolean.
Two-party key-exchange protocols aim to share a secret key between the parties, so correctness can informally be thought as the equality $k_A = k_B$, where $k_A$ and $k_B$ are the keys reconstructed by $A$ and $B$ respectively, as the main property after running the protocol.
\begin{lstlisting}[language=XML]
<correctness>
  <application function="eq">
    <argument id="kA"></argument>
    <argument id="kB"></argument>
  </application>
</correctness>
\end{lstlisting}
This can be extended to multi-party settings.

Considering the ProVerif plugin introduced in Section~\ref{sec:proverif}, one way to model correctness is as property over execution traces when attackers are passive and do not tamper with the messages.
A trace $t$ is a list that includes: sent and received messages, insertions to (private) tables, and events that must be explicitly recorded by the processes.
Events annotate processes marking important stages reached by the protocol but do not otherwise affect their behaviour.
Traces are analysed by the reasoning core of ProVerif and are determined by each possible ramification of concurrently executing processes.
A security property over execution traces is defined as a predicate over elements in a subset of the space of all traces $T$.
We base our definition of correctness on the event $c \in \intset_p^2 -> E$, as a pair of keys mapped to the space of events $E$.
Each argument of the event is stored by a different entity; so if the execution of the protocol records the event $c(k_A,k_B)$ with $k_A = k_B$ into a trace $t$ among the (infinite) traces $T$, then the two parties must have successfully exchanged the same key.
The existence of $c \in t$ is sufficient to prove correctness, as the processes are run only once and the adversary cannot inject messages (so $c$ cannot happen by chance).
More formally, we model {\em correctness} as
\begin{equation*}
  \forall k_A, k_B \in \intset_p. \exists c \in \intset_p^2, t \in T. c = (k_A,k_B) \land k_A = k_B \land c \in t.
\end{equation*}
Our plugin reflects it in programming code by
\begin{itemize}
  \item setting the adversary to passive; {\tt attacker = passive};
  \item creating two tables {\tt finalA(Zp)} and {\tt finalB(Zp)};
  \item injecting into the parties' processes a command to fill the tables with the shared key: {\tt insert finalA(exp(gy, x))} for Alice and analogously for Bob;
  \item appending a process {\tt agreement}, that can generate the desired event, to the protocol process:
    \begin{lstlisting}[language=None]
let agreement = get finalA(kA) in
                get finalB(kB) in
                event correctness(kA, kB).
    \end{lstlisting}
  \item and finally, creating the event {\tt event correctness(Zp, Zp)} and querying for it to be triggered by any execution where the keys are the same:
    \begin{lstlisting}[basicstyle=\ttfamily\small]
query k: Zp; event(correctness(k, k)).
    \end{lstlisting}
\end{itemize}
Briefly, ProVerif analyses the execution traces and returns whether or not the required event can be found in a trace.

\subsection{Diffie-Hellman key exchange}
\label{sec:dhke-security}
\begin{figure}[!ht]
  \begin{subfigure}[b]{\columnwidth}
    \begin{center}
      \includegraphics[width=\columnwidth]{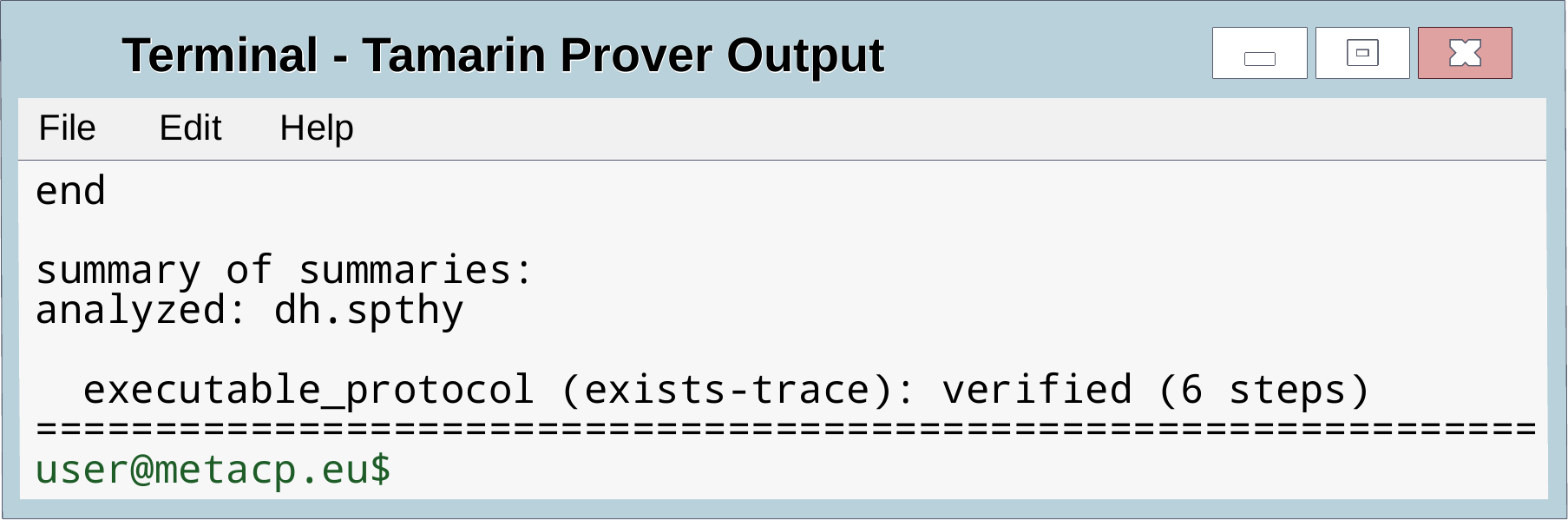}
    \end{center}
    \caption{Final part of Tamarin's output showing executability.}
    \Description{Execution of Tamarin for the executability of the protocol.}
    \label{fig:tamarin-output}
  \end{subfigure}
  \begin{subfigure}[b]{\columnwidth}
    \begin{center}
      \includegraphics[width=\columnwidth]{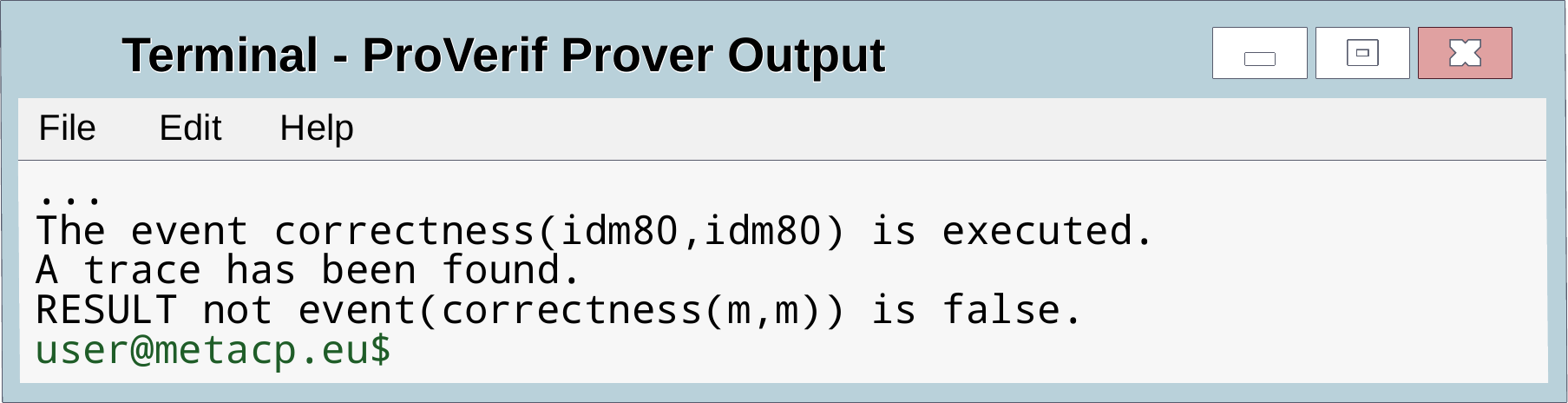}
    \end{center}
    \caption{Final part of ProVerif's output showing correctness.}
    \Description{Execution of ProVerif for the correctness property. If a trace is found, then correctness holds.}
    \label{fig:proverif-output}
  \end{subfigure}
  \caption{Security properties as formally verified in Tamarin (above) and in ProVerif (below).}
  \Description{Both executability and correctness are verified: the former in Tamarin, the latter in ProVerif.}
  \label{fig:proverif-tamarin-output}
\end{figure}

To have a complete picture of MetaCP, we also briefly illustrate the formal verification of the Diffie-Hellman key exchange to complete the steps described in Section~\ref{sec:dhke-model}.
The output from ProVerif running the automatically exported protocol, see Figure~\ref{fig:proverif-output}, shows the formal verification of correctness.
When the same specification is interpreted by the Tamarin plugins it, verifies the executability of the protocol, see Figure~\ref{fig:tamarin-output}.
We stress that this part is delegated to external tools that can parse the language exported through related MetaCP plugins, in our example ProVerif.

 \section{Conclusion and Future Work}

Motivated from the current limitations in manual interpretations to single tools, we explore the possibility of automatically generating multiple interpretations of cryptographic protocols from a single structured specification.
In particular, we demonstrate that the the expressiveness offered by this data-centric approach is enough to interpret a traditional key-exchange protocol into multiple (very different) formal verification languages and a popular object-oriented programming language.
In detail, this paper presented a new plugin for ProVerif, over the only plugin for Tamarin that existed before~\cite{arnaboldi2019poster}, and a new plugin to generate C++ code, that compiles and can run the protocol between devices over the Internet.
To do that, we delegated almost all semantic aspects to interpreting plugins; further research is required to understand how much semantics can be included in the specification language.
This is an important aspect toward the goals of formal verification, as, ultimately, the trustworthiness of verification results will either depend on the correctness of the interpretation, or, as at the current stage of development, require an expert in the target verification language.

Documentation and a demo of the tool have been released\footnote{Documentation and demo at \url{http://demo.metacp.eu}.
To reproduce our results, we share our implementation with instructions at \url{https://doi.org/10.5281/zenodo.6394688}.}.
The graphical interface does not (currently) support all the features provided by the specification, e.g., it only allows to design two-party protocols and no security properties can be specified yet.

After its promising entrance, MetaCP is far from a mature solution.
Future extensions envision the inclusions of more security properties (executability and correctness are not enough) along with the protocol model to further accelerate the verification process.
Furthermore, both ProVerif and Tamarin reason in the symbolic model -- we have yet to research into the translation targeting tools that reason in the computational model, e.g., CryptoVerif or EasyCrypt.
Having showcased the flexibility of the approach through these first plugins, further plugins are an obvious next step to incorporate further verification and programming languages alike.
Finally, our tool can be used to bring the benefits of formal verification to domains where security flaws can affect critical infrastructures, e.g., the electric vehicle charging infrastructure~\cite{metere2021securing} or smart building with distributed energy generation~\cite{vahidinasab2021active}.
This necessity has already been appreciated by Lauser et al.~\cite{lauser2020security} in the automotive industry.
We can see how a future extension of our approach could alleviate the security burden in the current design process for practitioners in a myriad of further security-critical domains.

\begin{acks}
This research was partly funded by The Alan Turing Institute
through the Lloyd's Register Foundation (G0095), an Innovate UK e4Future grant (104227), and EPSRC grants EP/S016627/1, Active Building Centre, and EP/T027037/1 AISEC.
\end{acks}

\bibliographystyle{ACM-Reference-Format}
\bibliography{references}


\begin{thebibliography}{21}


\ifx \showCODEN    \undefined \def \showCODEN     #1{\unskip}     \fi
\ifx \showDOI      \undefined \def \showDOI       #1{#1}\fi
\ifx \showISBNx    \undefined \def \showISBNx     #1{\unskip}     \fi
\ifx \showISBNxiii \undefined \def \showISBNxiii  #1{\unskip}     \fi
\ifx \showISSN     \undefined \def \showISSN      #1{\unskip}     \fi
\ifx \showLCCN     \undefined \def \showLCCN      #1{\unskip}     \fi
\ifx \shownote     \undefined \def \shownote      #1{#1}          \fi
\ifx \showarticletitle \undefined \def \showarticletitle #1{#1}   \fi
\ifx \showURL      \undefined \def \showURL       {\relax}        \fi
\providecommand\bibfield[2]{#2}
\providecommand\bibinfo[2]{#2}
\providecommand\natexlab[1]{#1}
\providecommand\showeprint[2][]{arXiv:#2}

\bibitem[\protect\citeauthoryear{Abadi, Blanchet, and Fournet}{Abadi
  et~al\mbox{.}}{2017}]%
        {abadi2017applied}
\bibfield{author}{\bibinfo{person}{Mart{\'\i}n Abadi}, \bibinfo{person}{Bruno
  Blanchet}, {and} \bibinfo{person}{C{\'e}dric Fournet}.}
  \bibinfo{year}{2017}\natexlab{}.
\newblock \showarticletitle{{The applied pi calculus: Mobile values, new names,
  and secure communication}}.
\newblock \bibinfo{journal}{\emph{Journal of the ACM (JACM)}}
  \bibinfo{volume}{65}, \bibinfo{number}{1} (\bibinfo{year}{2017}),
  \bibinfo{pages}{1--41}.
\newblock


\bibitem[\protect\citeauthoryear{Armando, Arsac, Avanesov, Barletta, Calvi,
  Cappai, Carbone, Chevalier, Compagna, Cu{\'e}llar, et~al\mbox{.}}{Armando
  et~al\mbox{.}}{2012}]%
        {armando2012avantssar}
\bibfield{author}{\bibinfo{person}{Alessandro Armando}, \bibinfo{person}{Wihem
  Arsac}, \bibinfo{person}{Tigran Avanesov}, \bibinfo{person}{Michele
  Barletta}, \bibinfo{person}{Alberto Calvi}, \bibinfo{person}{Alessandro
  Cappai}, \bibinfo{person}{Roberto Carbone}, \bibinfo{person}{Yannick
  Chevalier}, \bibinfo{person}{Luca Compagna}, \bibinfo{person}{Jorge
  Cu{\'e}llar}, {et~al\mbox{.}}} \bibinfo{year}{2012}\natexlab{}.
\newblock \showarticletitle{{The AVANTSSAR platform for the automated
  validation of trust and security of service-oriented architectures}}. In
  \bibinfo{booktitle}{\emph{International Conference on Tools and Algorithms
  for the Construction and Analysis of Systems}}. Springer,
  \bibinfo{pages}{267--282}.
\newblock


\bibitem[\protect\citeauthoryear{Armando, Basin, et~al\mbox{.}}{Armando
  et~al\mbox{.}}{2005}]%
        {armando2005avispa}
\bibfield{author}{\bibinfo{person}{Alessandro Armando}, \bibinfo{person}{David
  Basin}, {et~al\mbox{.}}} \bibinfo{year}{2005}\natexlab{}.
\newblock \showarticletitle{{The AVISPA tool for the Automated Validation of
  Internet Security Protocols and Applications}}. In
  \bibinfo{booktitle}{\emph{{International conference on Computer Aided
  Verification}}}. Springer, \bibinfo{pages}{281--285}.
\newblock


\bibitem[\protect\citeauthoryear{Arnaboldi and Metere}{Arnaboldi and
  Metere}{2019}]%
        {arnaboldi2019poster}
\bibfield{author}{\bibinfo{person}{Luca Arnaboldi} {and}
  \bibinfo{person}{Roberto Metere}.} \bibinfo{year}{2019}\natexlab{}.
\newblock \showarticletitle{{Poster: Towards a Data Centric Approach for the
  Design and Verification of Cryptographic Protocols}}. In
  \bibinfo{booktitle}{\emph{Proceedings of the 2019 ACM SIGSAC Conference on
  Computer and Communications Security}}. \bibinfo{pages}{2585--2587}.
\newblock


\bibitem[\protect\citeauthoryear{Barthe, Gr{\'e}goire, Heraud, and
  B{\'e}guelin}{Barthe et~al\mbox{.}}{2011}]%
        {barthe2011computer}
\bibfield{author}{\bibinfo{person}{Gilles Barthe}, \bibinfo{person}{Benjamin
  Gr{\'e}goire}, \bibinfo{person}{Sylvain Heraud}, {and}
  \bibinfo{person}{Santiago~Zanella B{\'e}guelin}.}
  \bibinfo{year}{2011}\natexlab{}.
\newblock \showarticletitle{{Computer-aided security proofs for the working
  cryptographer}}. In \bibinfo{booktitle}{\emph{Annual Cryptology Conference}}.
  Springer, \bibinfo{pages}{71--90}.
\newblock


\bibitem[\protect\citeauthoryear{Basin, Dreier, Hirschi, Radomirovic, Sasse,
  and Stettler}{Basin et~al\mbox{.}}{2018}]%
        {basin2018formal}
\bibfield{author}{\bibinfo{person}{David Basin}, \bibinfo{person}{Jannik
  Dreier}, \bibinfo{person}{Lucca Hirschi}, \bibinfo{person}{Sa{\v{s}}a
  Radomirovic}, \bibinfo{person}{Ralf Sasse}, {and} \bibinfo{person}{Vincent
  Stettler}.} \bibinfo{year}{2018}\natexlab{}.
\newblock \showarticletitle{{A formal analysis of 5G authentication}}. In
  \bibinfo{booktitle}{\emph{Proceedings of the 2018 ACM SIGSAC conference on
  computer and communications security}}. \bibinfo{pages}{1383--1396}.
\newblock


\bibitem[\protect\citeauthoryear{Blanchet}{Blanchet}{2016}]%
        {blanchet2016modeling}
\bibfield{author}{\bibinfo{person}{Bruno Blanchet}.}
  \bibinfo{year}{2016}\natexlab{}.
\newblock \showarticletitle{{Modeling and verifying security protocols with the
  applied pi calculus and ProVerif}}.
\newblock \bibinfo{journal}{\emph{Foundations and Trends{\textregistered} in
  Privacy and Security}} \bibinfo{volume}{1}, \bibinfo{number}{1-2}
  (\bibinfo{year}{2016}), \bibinfo{pages}{1--135}.
\newblock


\bibitem[\protect\citeauthoryear{Blanchet et~al\mbox{.}}{Blanchet
  et~al\mbox{.}}{2001}]%
        {blanchet2001efficient}
\bibfield{author}{\bibinfo{person}{Bruno Blanchet} {et~al\mbox{.}}}
  \bibinfo{year}{2001}\natexlab{}.
\newblock \showarticletitle{{An Efficient Cryptographic Protocol Verifier Based
  on Prolog Rules.}}. In \bibinfo{booktitle}{\emph{csfw}},
  Vol.~\bibinfo{volume}{1}. \bibinfo{pages}{82--96}.
\newblock


\bibitem[\protect\citeauthoryear{Clark, DeRose, et~al\mbox{.}}{Clark
  et~al\mbox{.}}{1999}]%
        {clark1999xml}
\bibfield{author}{\bibinfo{person}{James Clark}, \bibinfo{person}{Steve
  DeRose}, {et~al\mbox{.}}} \bibinfo{year}{1999}\natexlab{}.
\newblock \bibinfo{title}{{XML path language (XPath) version 1.0}}.
\newblock
\newblock


\bibitem[\protect\citeauthoryear{Cremers and Horvat}{Cremers and
  Horvat}{2016}]%
        {cremers2016improving}
\bibfield{author}{\bibinfo{person}{Cas Cremers} {and} \bibinfo{person}{Marko
  Horvat}.} \bibinfo{year}{2016}\natexlab{}.
\newblock \showarticletitle{{Improving the ISO/IEC 11770 standard for key
  management techniques}}.
\newblock \bibinfo{journal}{\emph{International Journal of Information
  Security}} \bibinfo{volume}{15}, \bibinfo{number}{6} (\bibinfo{year}{2016}),
  \bibinfo{pages}{659--673}.
\newblock


\bibitem[\protect\citeauthoryear{Dai}{Dai}{[n.d.]}]%
        {cryptopp}
\bibfield{author}{\bibinfo{person}{Wei Dai}.}
  \bibinfo{year}{[n.d.]}\natexlab{}.
\newblock \bibinfo{title}{{Crypto++® library}}.
\newblock
\newblock
\urldef\tempurl%
\url{https://cryptopp.com/}
\showURL{%
\tempurl}


\bibitem[\protect\citeauthoryear{Denker and Millen}{Denker and Millen}{2000}]%
        {denker2000capsl}
\bibfield{author}{\bibinfo{person}{Grit Denker} {and} \bibinfo{person}{Jonathan
  Millen}.} \bibinfo{year}{2000}\natexlab{}.
\newblock \showarticletitle{{CAPSL integrated protocol environment}}. In
  \bibinfo{booktitle}{\emph{Proceedings DARPA Information Survivability
  Conference and Exposition. DISCEX'00}}, Vol.~\bibinfo{volume}{1}. IEEE,
  \bibinfo{pages}{207--221}.
\newblock


\bibitem[\protect\citeauthoryear{{Hao, Feng and Metere, Roberto and
  Shahandashti, Siamak F and Dong, Changyu}}{{Hao, Feng and Metere, Roberto and
  Shahandashti, Siamak F and Dong, Changyu}}{2018}]%
        {hao2018speke}
\bibfield{author}{\bibinfo{person}{{Hao, Feng and Metere, Roberto and
  Shahandashti, Siamak F and Dong, Changyu}}.} \bibinfo{year}{2018}\natexlab{}.
\newblock \showarticletitle{{Analyzing and patching SPEKE in ISO/IEC}}.
\newblock \bibinfo{journal}{\emph{{IEEE Transactions on Information Forensics
  and Security}}} \bibinfo{volume}{13}, \bibinfo{number}{11}
  (\bibinfo{year}{2018}), \bibinfo{pages}{2844--2855}.
\newblock


\bibitem[\protect\citeauthoryear{{Kobeissi, Nadim and Bhargavan, Karthikeyan
  and Blanchet, Bruno}}{{Kobeissi, Nadim and Bhargavan, Karthikeyan and
  Blanchet, Bruno}}{2017}]%
        {kobeissi2017automated}
\bibfield{author}{\bibinfo{person}{{Kobeissi, Nadim and Bhargavan, Karthikeyan
  and Blanchet, Bruno}}.} \bibinfo{year}{2017}\natexlab{}.
\newblock \showarticletitle{{Automated verification for secure messaging
  protocols and their implementations: A symbolic and computational approach}}.
  In \bibinfo{booktitle}{\emph{{IEEE EuroS\&P}}}.
\newblock


\bibitem[\protect\citeauthoryear{Lauser, Zelle, and Krau{\ss}}{Lauser
  et~al\mbox{.}}{2020}]%
        {lauser2020security}
\bibfield{author}{\bibinfo{person}{Timm Lauser}, \bibinfo{person}{Daniel
  Zelle}, {and} \bibinfo{person}{Christoph Krau{\ss}}.}
  \bibinfo{year}{2020}\natexlab{}.
\newblock \showarticletitle{{Security Analysis of Automotive Protocols}}. In
  \bibinfo{booktitle}{\emph{{Computer Science in Cars Symposium}}}.
  \bibinfo{pages}{1--12}.
\newblock


\bibitem[\protect\citeauthoryear{Meier, Schmidt, Cremers, and Basin}{Meier
  et~al\mbox{.}}{2013}]%
        {meier2013tamarin}
\bibfield{author}{\bibinfo{person}{Simon Meier}, \bibinfo{person}{Benedikt
  Schmidt}, \bibinfo{person}{Cas Cremers}, {and} \bibinfo{person}{David
  Basin}.} \bibinfo{year}{2013}\natexlab{}.
\newblock \showarticletitle{{The TAMARIN prover for the symbolic analysis of
  security protocols}}. In \bibinfo{booktitle}{\emph{International Conference
  on Computer Aided Verification}}. Springer, \bibinfo{pages}{696--701}.
\newblock


\bibitem[\protect\citeauthoryear{Metere, Neaimeh, Morisset, Maple, Bellekens,
  and Czekster}{Metere et~al\mbox{.}}{2021}]%
        {metere2021securing}
\bibfield{author}{\bibinfo{person}{Roberto Metere}, \bibinfo{person}{Myriam
  Neaimeh}, \bibinfo{person}{Charles Morisset}, \bibinfo{person}{Carsten
  Maple}, \bibinfo{person}{Xavier Bellekens}, {and} \bibinfo{person}{Ricardo~M
  Czekster}.} \bibinfo{year}{2021}\natexlab{}.
\newblock \showarticletitle{{Securing the Electric Vehicle Charging
  Infrastructure}}.
\newblock \bibinfo{journal}{\emph{arXiv preprint arXiv:2105.02905}}
  (\bibinfo{year}{2021}).
\newblock


\bibitem[\protect\citeauthoryear{ThinkAsynch}{ThinkAsynch}{[n.d.]}]%
        {asiocpp}
\bibfield{author}{\bibinfo{person}{ThinkAsynch}.}
  \bibinfo{year}{[n.d.]}\natexlab{}.
\newblock \bibinfo{title}{{Asio C++ Library}}.
\newblock
\newblock
\urldef\tempurl%
\url{https://think-async.com/Asio/}
\showURL{%
\tempurl}


\bibitem[\protect\citeauthoryear{Thompson, Mendelsohn, Beech, and
  Maloney}{Thompson et~al\mbox{.}}{2009}]%
        {thompson2009w3c}
\bibfield{author}{\bibinfo{person}{Henry~S Thompson}, \bibinfo{person}{Noah
  Mendelsohn}, \bibinfo{person}{D Beech}, {and} \bibinfo{person}{M Maloney}.}
  \bibinfo{year}{2009}\natexlab{}.
\newblock \showarticletitle{{W3C XML schema definition language (XSD) 1.1 part
  1: Structures}}.
\newblock \bibinfo{journal}{\emph{The World Wide Web Consortium (W3C), W3C
  Working Draft Dec}}  \bibinfo{volume}{3} (\bibinfo{year}{2009}).
\newblock


\bibitem[\protect\citeauthoryear{Vahidinasab, Ardalan, Mohammadi-Ivatloo,
  Giaouris, and Walker}{Vahidinasab et~al\mbox{.}}{2021}]%
        {vahidinasab2021active}
\bibfield{author}{\bibinfo{person}{Vahid Vahidinasab}, \bibinfo{person}{Chenour
  Ardalan}, \bibinfo{person}{Behnam Mohammadi-Ivatloo}, \bibinfo{person}{Damian
  Giaouris}, {and} \bibinfo{person}{Sara~L Walker}.}
  \bibinfo{year}{2021}\natexlab{}.
\newblock \showarticletitle{{Active building as an energy system: concept,
  challenges, and outlook}}.
\newblock \bibinfo{journal}{\emph{IEEE Access}}  \bibinfo{volume}{9}
  (\bibinfo{year}{2021}), \bibinfo{pages}{58009--58024}.
\newblock


\bibitem[\protect\citeauthoryear{Vigan{\`o}}{Vigan{\`o}}{2006}]%
        {vigano2006automated}
\bibfield{author}{\bibinfo{person}{Luca Vigan{\`o}}.}
  \bibinfo{year}{2006}\natexlab{}.
\newblock \showarticletitle{{Automated security protocol analysis with the
  {AVISPA} tool}}.
\newblock \bibinfo{journal}{\emph{Electronic Notes in Theoretical Computer
  Science}}  \bibinfo{volume}{155} (\bibinfo{year}{2006}),
  \bibinfo{pages}{61--86}.
\newblock


\end{thebibliography}

\end{document}